\documentstyle[aps,prb,multicol,amsfonts,epsf]{revtex}

\begin{document}
\draft
\title{Corrections to the universal behavior of the Coulomb-blockade 
peak splitting for quantum dots separated by a 
finite barrier}
\author{John M. Golden and Bertrand I. Halperin}
\address{Department of Physics, Harvard University, Cambridge, MA
  02138}

\date{21 November 1996}

\maketitle

\begin{abstract}
\noindent \ \
Building upon earlier work on the
relation between the dimensionless interdot
channel conductance $g$ and the fractional
Coulomb-blockade peak splitting $f$ for two 
electrostatically equivalent dots,
we calculate the leading correction that
results from an interdot tunneling barrier
that is not a delta-function but, rather,
has a finite height $V_{0}$ and a 
nonzero width $\xi$ and can be approximated
as parabolic near its peak.  We develop a
new treatment of the problem for $g \ll 1$ 
that starts from the single-particle eigenstates 
for the full coupled-dot system.
The finiteness of the barrier
leads to a small upward shift of the
$f$-versus-$g$ curve for $g \ll 1$.
The shift is a consequence of the fact that
the tunneling matrix elements vary exponentially
with the energies of the states connected.
Therefore, when $g$ is small, it can pay
to tunnel to intermediate
states with single-particle energies above 
the barrier height $V_{0}$.  For a parabolic barrier, 
the energy scale for the variation in the tunneling
matrix elements is $\hbar \omega$, where $\omega$, 
which is proportional to $\sqrt{V_{0}}/\xi$, is 
the harmonic oscillator frequency of the inverted
parabolic well.  The size of the correction to
previous zero-width ($\xi = 0$) calculations
depends strongly on the ratio between 
$\hbar \omega$ and the energy cost $U$ 
associated with moving electrons between the dots.
In the limit $g \rightarrow 0$, the finite-width
$f$-versus-$g$ curve behaves like 
$(U/\hbar \omega)/|\ln g|$.  
The correction to the zero-width behavior
does not affect agreement with recent experiments
in which $2\pi U/\hbar \omega \simeq  1$ but
may be important in future experiments.
\end{abstract}


\begin{multicols}{2}

\narrowtext

\section{Introduction}

The opening of tunneling channels between two
quantum dots leads to a transition from a Coulomb
blockade characteristic of isolated dots to
one characteristic of a single large composite
dot.~\cite{Review}  For a pair of 
electrostatically identical quantum dots,
this transformation can be chronicled by tracking the 
splitting of
the Coulomb blockade conductance peaks as a function of
the conductance through the interdot tunneling
channels.~\cite{Waugh,Crouch,Livermore}
If one assumes a single common value for the conductance
in each tunneling channel (an assumption that is
exactly fulfilled for a spin-symmetric system of only 
two channels, one for spin-up electrons and the other for 
spin-down electrons), one can divide the peak splitting 
by its saturation value and look for the relation between
two dimensionless quantities:~\cite{Golden1,Golden2}
the fractional peak splitting $f$ and the 
dimensionless channel conductance $g$.~\cite{Comment1}  

For $g \ll 1$,
previous theoretical work~\cite{Golden1,Golden2,Matveev3,%
Matveev4} has treated the coupled-dot problem
via a ``transfer-Hamiltonian approach,''~\cite{Duke} in
which states localized on one dot are connected to those
localized on the other by hopping matrix elements.
(Here {\it localized} signifies that a
state is entirely restricted to one of the two dots.)
The hopping matrix elements have been treated as constant, 
independent of the states connected, as they would be
if the interdot barrier were a delta-function potential,
having infinite height and zero width.
For such a barrier, the leading small-$g$ behavior of 
the fractional peak splitting is given by
\begin{equation}
f_{\xi = 0}^{(1)} = \frac{2 \ln 2}{\pi^2} N_{\text{ch}} g,
\label{eq:fzero}
\end{equation}
where $N_{\text{ch}}$ is the number
of separate tunneling modes
(spin-up and spin-down channels are counted separately).
The superscript of $f_{\xi = 0}^{(1)}$ tells us that this 
is the leading term in the weak-coupling limit.  The
subscript further specifies that this term is calculated
for a tunneling barrier of effectively zero width
$(\xi = 0)$ and therefore, by implication, of infinite height.
For $g \gtrsim 0.2$, subleading terms, which
are higher-order in $g$, contribute significantly to the
zero-width peak splitting, and the
first set of these, which consists of terms
proportional to $g^2$, has been calculated in 
previous work.~\cite{Golden2}  

In this paper, we calculate a different correction
to the $\xi = 0$, first-order in $g$ result which 
arises from the 
fact that a realistic barrier possesses a finite
height $V_{0}$ and a nonzero width $\xi$.  
For such a realistic barrier,
the hopping matrix elements that move electrons between
the dots are not independent of the states they connect, 
and, for small $g$, they depend exponentially on the
energies of the states.  
As a result of this exponential dependence, in the
weak-coupling ($g \ll 1$) limit, it can 
pay to tunnel to intermediate states with 
energies above the barrier, and the leading term
in the fractional peak splitting then behaves as
$(U/W)/|\ln g|$, where
$U$ is the interdot charging energy, which
measures the capacitive energy cost of moving electrons 
between the dots, and $W$ is the characteristic
energy scale over which the hopping matrix 
elements change from their values at the Fermi energy
$E_{\text{F}}$. 

We examine specifically the case 
of a finite-width interdot barrier that can be 
treated as parabolic near its peak.
We find that, for such a barrier, the energy scale $W$ is
equal to $\hbar \omega/2 \pi$, where $\omega$
is the harmonic oscillator

\begin{minipage}{3.27truein}
  \begin{figure}[H]
    \begin{center}
      \leavevmode
      \epsfxsize=3.27truein
      \epsfbox{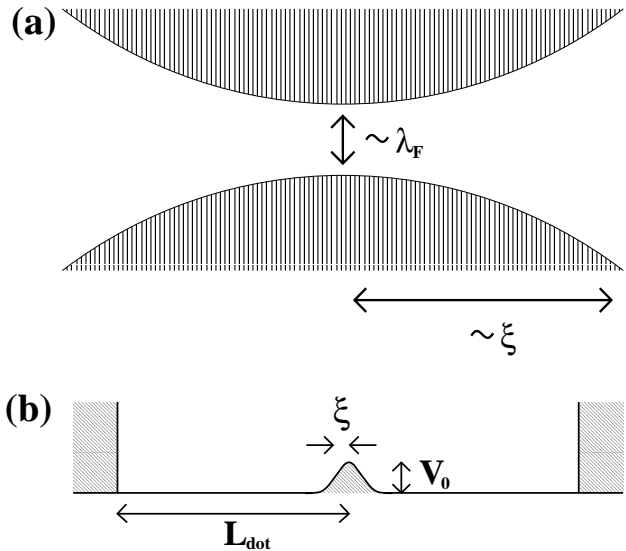}
    \end{center}
    \caption{(a) Schematic diagram for a single orbital-mode 
connection between the two dots.  Over 
a distance of order $\xi$, the connection
narrows to a minimum width on the order of the Fermi
wavelength $\lambda_{\text{F}}$.
(b) ``Box-like'' double-dot system with a central
barrier.  Hard confining walls are located at a distance
$L_{\text{dot}}$ from the barrier.  The barrier is
characterized by its height $V_{0}$ and half width $\xi$.}
    \label{fig:barrier}
  \end{figure}
  \smallskip
\end{minipage}

\noindent frequency of the
inverted parabolic well.  This frequency is 
proportional to the square root of the
barrier curvature, which is proportional
to $V_{0}/\xi^2$.  It follows that the limit
$\xi \rightarrow 0$ corresponds to the limit
$U/W \rightarrow 0$.  For $\xi \neq 0$, on the other hand,
it is not always true that $U/W \ll 1$.  In fact, in
recent experiments by Waugh {\em et al.}~\cite{Waugh},
Crouch {\em et al.}~\cite{Crouch},
and Livermore {\it et al.}~\cite{Livermore}, 
it appears that $U/W$ is roughly $1$.  Under
such circumstances, we find that, for a given small value of 
the channel conductance ($g \ll 1$), the fractional
peak splitting $f$ is larger than the
zero-width splitting, $f_{\xi = 0}$, by a small but
noticeable amount, and, in the extreme limit of 
$g \rightarrow 0$, the ratio of the finite-width 
peak splitting to the previously calculated zero-width
peak splitting becomes very large.
For intermediate values of $g$, on the other hand, the
primary effect is a small increase in $f$ accompanied by
a reduction of the slope of the $f$-versus-$g$ curve.  

To find the leading term in the finite-width fractional
peak splitting we adopt 
a {\it stationary-state approach},~\cite{Duke} in which
the first step is to solve for the single-particle 
eigenstates of non-interacting electrons moving in 
the electrostatic potential of the coupled dots.
The capacitive 
interactions between the electrons  are then expressed
in terms of these
non-interacting double-dot eigenstates, and the off-diagonal
elements of the interactions are treated perturbatively.
The leading term in the finite-width
fractional peak splitting, $f^{(1)}$, is determined by
finding the value for $\rho = 1$ of a 
more general quantity 
$\tilde{f}^{(1)}(\rho)$, where $\rho$ is a dimensionless
parameter (defined by Eq.~\ref{eq:HC1} below) which is a measure
of the bias asymmetry between the dots.~\cite{Golden1,Golden2}
In the limit $U/W \rightarrow 0$, the zero-width result, 
$f_{\xi = 0}^{(1)}$, is recovered.
For finite $U/W$, an approximate analytic calculation   
demonstrates the limiting $1/|\ln g|$ behavior, which
is confirmed by numerical results.
For the particular choice $U/W = 1$, which  
corresponds to recent experiments,~\cite{Waugh,Crouch,Livermore}
as well as for various other choices of the ratio $U/W$,
the leading term in the fractional peak splitting is computed
numerically as a function of $g$.  It is confirmed that 
the condition $\xi \neq 0$ leads to
an upward shift of the peak splitting for
weakly coupled dots ($g \ll 1$).
As $g$ becomes larger, the effect
of allowing $\xi \neq 0$ becomes less dramatic, and,
for $U/W \simeq 1$,
previous predictions for the fractional peak splitting
at intermediate values of $g$ are essentially unaltered.  

The structure of this paper is as follows.  
Sec.~II develops the stationary-state approach for 
calculating $\tilde{f}(\rho)$.  Sec.~III implements
this approach for a parabolic interdot barrier,
verifying the $1/|\ln g|$ behavior of the $g \rightarrow 0$
peak splitting that arises for $\xi \neq 0$ and
putting the finite-width calculation in the context
of earlier work.
Sec.~IV summarizes the results and comments on the
possible effects of $\xi \neq 0$ when the dots are
strongly coupled ($g \simeq 1$).

\section{The Stationary-State Approach}

In order to solve for $\tilde{f}(\rho)$ via
the stationary-state approach, we make the problem
one-dimensional by considering a smooth, adiabatic
interdot connection [see Fig.~1(a)] which, for simplicity,
we presume to contain only one transverse
orbital mode that lies near or below the Fermi 
energy $E_{\text{F}}$.~\cite{Matveev2}  (The use of one orbital mode
corresponds to the spin-symmetric $N_{\text{ch}} = 2$
experiments of Waugh {\it et al.}~\cite{Waugh}, 
Crouch {\it et al.}~\cite{Crouch},
and Livermore {\it et al.}~\cite{Livermore})
For such a single orbital-mode 
connection, the only parts of an electron
wavefunction that can pass from dot to dot are those
that overlap with the lowest transverse mode.  Hence, in
investigating the effect of the interdot connection,
we can ignore all electrons but those in this lowest 
mode.  We are left with a one-dimensional problem in which 
a representative electron with single-particle energy
$E$ moves in an effective potential 
$V(x) = E_{\text{tr}}(x) + V_{\text{el}}(x)$,
where $E_{\text{tr}}(x)$ is the spatially dependent
energy of the lowest transverse mode and $V_{\text{el}}(x)$
is the spatially dependent electrostatic energy.
The characteristic length scale for the 
spatial variation of the effective potential is
the barrier width $\xi$. 

After adding hard boundaries 
at a distance $L_{\text{dot}}$ from the barrier, we have a
``box-like'' double-dot system
with a smoothly varying longitudinal potential [see Fig.~1(b)].
The Hamiltonian consists of two components.
The first, $H_{0}$, is a diagonal term 
giving the energies of the non-interacting, single-particle
eigenstates, which form a discrete spectrum
with an average level spacing proportional to 
$\hbar v_{\text{F}}/L_{\text{dot}}$ near the Fermi surface,
where $v_{\text{F}}$ is the Fermi velocity.
The second, $H_{C}$, gives the
capacitive energy cost of moving electrons
from one side of the barrier to the other:~\cite{Golden1,%
Golden2}
\begin{equation}
H_{C} = U (\hat{n} - \rho/2)^{2} \, .
\label{eq:HC1}
\end{equation}
Here, $\hat{n}$ counts 
the electrons transferred from dot 1 to dot 2
(assuming, for convenience, an even total number of electrons 
initially divided equally between the two dots).  The 
dimensionless parameter 
$\rho$ is a measure of the capacitively weighted bias
and favors occupation of dot 2 when 
$\rho > 0$.  (Note that
$U$ equals the quantity $U_{2}$ of previous 
work.~\cite{Golden1,Golden2})

We need to solve for the dimensionless channel conductance
$g$ and the fractional peak splitting $f$.
For the non-interacting electrons characterized by $H_{0}$,
the dimensionless channel conductance $g$ 
is simply the transmission
probability for a particle incident on the barrier 
at the Fermi energy $E_{\text{F}}$.~\cite{Duke}  
$f$ is not so easily determined because, in its evaluation,
$H_{C}$ is relevant.  Thus,  
we must develop a means of dealing with $\hat{n}$,
which is not diagonal in the basis of non-interacting
single-particle eigenstates that is the cornerstone
of our approach.

Our strategy is
to switch to a basis that is simply related 
to the eigenstate basis but which renders $\hat{n}$
nearly diagonal at energies that are low compared to
the barrier.  We use the fact that, for a bound system
containing two equal potential minima, the
eigenstates come in well-defined, discrete pairs.~\cite{Hund}  
The states in these pairs have similar energies 
but opposite parities, the even-parity state having a
lower energy than the odd-parity state.  Thus, the
non-interacting part of the Hamiltonian can
be written in the form
\begin{equation}
H_{0} = \sum_{\sigma,j} E_{S}(j) 
            c^{\dagger}_{S j \sigma} \, c_{S j \sigma}
       +\sum_{\sigma,j} E_{A}(j) 
            c^{\dagger}_{A j \sigma} \, c_{A j \sigma} \, ,
\label{eq:evenodd}
\end{equation}
where $S$ and $A$ are the even and odd parity
indices, $j$ is the pair index, and $\sigma$ is the
spin index (which might be more generally regarded as a
channel index).

At lower and lower energies relative to the barrier,
the splitting within the pairs, $|E_{A}(j) - E_{S}(j)|$,
approaches zero, but the spacing
between pairs, $|E_{S}(j+1) - E_{A}(j)|$, remains approximately 
equal to $\delta$, where
$\delta = \pi \hbar v_{\text{F}}/L_{\text{dot}}$
(assuming we do not stray too far from the Fermi
surface).  It follows that, at low energies, one can form  
doublets of {\it quasi-localized} 
states---states that
lie mostly on one of the two sides of the central
barrier---from linear combinations of the symmetric
and antisymmetric components of each eigenstate pair.~\cite{Hund}
If $\phi_{S j}(x)$ and $\phi_{A j}(x)$ 
are the symmetric and antisymmetric eigenfunctions 
of the $j$th lowest-energy pair (with appropriately
chosen overall phases), the recipe for the
quasi-localized wavefunctions is
\begin{equation}
\Omega_{j \alpha}(x) = \frac{1}{\sqrt{2}} \left[
      \phi_{S j}(x) + (-1)^{\alpha + 1} \phi_{A j}(x) \right],
\label{eq:qlwave1}
\end{equation}
where the dot index $\alpha$ signifies that $\Omega_{j \alpha}(x)$
is primarily localized on the dot 1 side of the barrier
if $\alpha = 1$ and on the dot 2 side if $\alpha = 2$.     

%

At high energies relative to the barrier, we cannot form
quasi-localized wavefunctions from combinations of just
two states.
Nevertheless, we continue to form the linear combinations 
analogous to those of Eq.~\ref{eq:qlwave1}.
We refer to the full set of states $\Omega_{j \alpha}(x)$ 
as {\it semi-localized} to indicate
that these states are sometimes quasi-localized
(i.e., when they lie at low energies relative to
the interdot barrier) and sometimes not.  

The semi-localized states constitute the complete and
orthogonal basis that we
need to render $\hat{n}$ nearly diagonal
at low energies.
Their simple relation to the double-dot
eigenstates translates into an equally simple
relation between the corresponding creation and
annihilation operators.
The semi-localized annihilation
operators are given by
\begin{equation}
a_{j \alpha \sigma} = \frac{1}{\sqrt{2}}
 \left[ c_{S j \sigma} + (-1)^{\alpha + 1}
                    c_{A j \sigma} \right] \, , 
\label{eq:opLR}
\end{equation}
and the corresponding expression for $H_{0}$ is
\begin{equation}
H_{0} = \sum_{\sigma,\alpha,j} E(j) 
             a^{\dagger}_{j \alpha \sigma} a_{j \alpha \sigma}
    -\sum_{\sigma,j} t(j) 
          (a^{\dagger}_{j 2 \sigma} a_{j 1 \sigma} 
                              +  \text{H.c.}) \, ,
\label{eq:quasiloc}
\end{equation}
where  $E(j)$ is the average energy of the pair and
$t(j)$ is half the energy difference within the pair:
\begin{eqnarray}
E(j) &  = & \frac{E_{A}(j) + E_{S}(j)}{2} \, , \nonumber \\
t(j) & =  & \frac{E_{A}(j) - E_{S}(j)}{2} \, .
\label{eq:En&tn}
\end{eqnarray}
It is important to note that, whereas $E(j)$ is in 
general on the order of the Fermi energy, $t(j)$ is 
no greater than the average
level spacing $\delta = \pi \hbar v_{\text{F}}/L_{\text{dot}}$
and becomes vanishingly small in the large-dot
limit ($L_{\text{dot}} \rightarrow \infty$).  The minuteness
of $t(j)$ will permit us to ignore it in calculating the 
leading contribution to the fractional peak splitting.

We now write $\hat{n}$ in terms of the semi-localized operators.  
If dot 1 corresponds to the $x < 0$ side of the barrier
and dot 2 corresponds to the $x > 0$ side, we have
\begin{equation}
\hat{n} = \frac{1}{2} \int \! dx \left[\Theta(x)
                 - \Theta(-x) \right]
           \psi^{\dagger}(x) \psi(x),
\label{eq:nstep}
\end{equation}
where $\psi(x)$ is the position operator and 
$\Theta(x)$ is the Heaviside step function.~\cite{CommSmear} 

After writing $\psi(x)$ in terms of the semi-localized
operators $a_{j \alpha \sigma}$, we see that 
$\hat{n} = \hat{n}_{0} + \delta \hat{n}_{\text{C}} 
             + \delta \hat{n}_{\text{T}}$,
where $\hat{n}_{0}$ corresponds to the $\hat{n}$
we would have if the semi-localized states 
were truly localized, $\delta \hat{n}_{\text{C}}$ 
is the part of $\delta \hat{n} = (\hat{n} - \hat{n}_{0})$ 
that does not transfer electrons from dot 1 to dot 2,
and $\delta \hat{n}_{\text{T}}$ is the part of
$\delta \hat{n}$ that does effect such a transfer:
\begin{eqnarray}
\hat{n}_{0} & = & \sum_{\sigma,\alpha,j}
     \frac{(-1)^{\alpha}}{2}
     a^{\dagger}_{j \alpha \sigma} a_{j \alpha \sigma},
                 \nonumber \\
\delta \hat{n}_{\text{C}} & = &  
     \sum_{\sigma,\alpha,j_{1},j_{2}}
     \left[ B(j_{2},\alpha;j_{1},\alpha) 
      - \frac{(-1)^{\alpha}}{2} \delta_{j_{1},j_{2}} \right]
     a^{\dagger}_{j_{2} \alpha \sigma} a_{j_{1} \alpha \sigma},
                 \nonumber \\
\delta \hat{n}_{T} & = &  
\sum_{\sigma,\alpha,j_{1},j_{2}}
     B(j_{2},\bar{\alpha};j_{1},\alpha)   
     a^{\dagger}_{j_{2} \bar{\alpha} \sigma} 
        a_{j_{1} \alpha \sigma} \, .
\label{eq:nparts2}
\end{eqnarray}
Here, $\bar{\alpha}$ means ``not $\alpha$'' and
\begin{eqnarray}
B(j_{2},\alpha_{2};j_{1},\alpha_{1}) & = &
     \frac{1}{2} \int_{0}^{L_{\text{dot}}} \! dx \, \,
         \left[ (-1)^{\alpha_{1} + 1} \phi^{*}_{S j_{2}}(x) \,
                                 \phi_{A j_{1}}(x) \right.
   \nonumber \\
 & & \hspace{0.1in} \left. \mbox{} 
             + (-1)^{\alpha_{2} + 1} \phi^{*}_{A j_{2}}(x) \,
                                 \phi_{S j_{1}}(x) \right].
\label{eq:ncoeff2}
\end{eqnarray}

We have
obtained the desired ``semi-diagonal'' form of $\hat{n}$.
Using
$\delta \hat{n} = 
   \delta \hat{n}_{\text{C}} + \delta \hat{n}_{\text{T}}$
and assuming that $g$ is small,
we express the Hamiltonian in terms of
one non-perturbative piece, $H'_{0}$, and two perturbative
pieces, $H'_{\text{T}}$ and $H'_{\text{C}}$:
\begin{eqnarray}
H'_{0} & = & \sum_{\sigma,\alpha,j} E(j) 
                a^{\dagger}_{j \alpha \sigma} a_{j \alpha \sigma}
            + U(\hat{n}_{0} - \rho/2)^{2}, \nonumber \\
H'_{\text{T}} & = & -\sum_{\sigma,j} t(j) 
          (a^{\dagger}_{j  2 \sigma} a_{j 1 \sigma} 
                              +  \text{H.c.}), \nonumber \\
H'_{\text{C}} & = &  U(\hat{n}_{0} - \rho/2) \delta \hat{n}
                   + U \delta \hat{n} (\hat{n}_{0} - \rho/2)
                   + U (\delta \hat{n})^{2}.
\label{eq:Hprime}
\end{eqnarray}

As in Refs.~5 and 6, the
fractional peak splitting is determined
from $\tilde{f}(\rho)$, where
\begin{equation}
\tilde{f}(\rho) = \frac{\Delta(0) - \Delta(\rho)}{U/4}
\label{eq:frho}
\end{equation}
and $\Delta(\rho)$ is the energy shift of the ground state
of $H'_{0}$ due to the perturbations 
$H'_{\text{T}}$ and $H'_{\text{C}}$ for the given value of
$\rho$, where $0 \leq \rho < 1$ and the total number of
particles in the double-dot system is even.  
The quantity $\left[ \lim_{\rho \rightarrow 1} 
\tilde{f}(\rho) \right]$ 
equals the fractional peak splitting $f$.

Eq.~\ref{eq:frho} tells us that we are only interested
in relative energy shifts.  Consequently,
we can ignore terms such as 
$\langle 0 | U(\delta \hat{n})^2 | 0 \rangle$
that are independent of $\rho$.
(Here the brackets indicate an expectation value
taken in the ground state of $H'_{0}$.)
Another set of irrevelant terms are those of the
form $\langle 0 | U( \hat{n}_{0} - \rho/2) \delta \hat{n} | 0 \rangle$,
which are zero due to the symmetry of the ground
state with respect to interchange of the two dots.
Finally, terms that contain $H'_{\text{T}}$ are also negligible
because $t(j)$ goes to zero with the reciprocal of the
system size and, unlike $\delta \hat{n}$,
$H'_{\text{T}}$ only connects each state to one other, rather
than connecting each state to a manifold of others (see
Ref.~5 for a similar situation with regard
to odd orders in the transfer-Hamiltonian perturbation theory).
After the above terms are omitted, it is apparent
that the leading perturbative energy shift
comes from the term that is second order in $H'_{\text{C}}$.
To lowest order in $\delta \hat{n}$, this term is
\begin{equation}
\Delta^{(2)}(\rho) = -U^{2} \left\langle 0 \left| \delta \hat{n} 
  \, P_{0}  \frac{(\hat{n}_{0} - \rho)^{2}}{H'_{0} - E'_{0}(\rho)}
    P_{0} \, \delta \hat{n} \right| 0 \right\rangle,
\label{eq:Del1}
\end{equation}
where $E'_{0}(\rho)$ is the energy of the ground state of $H'_{0}$
and where $P_{0}$ is the operator that projects out the 
unperturbed ground state.  
$\Delta^{(2)}(\rho)$ can easily be seen to consist of two distinct
parts: a term second-order in $\delta \hat{n}_{C}$, which
involves hopping between states semi-localized on the same
dot, and a term second-order in $\delta \hat{n}_{T}$,
which involves hopping between states on different dots.

With Eq.~\ref{eq:Del1}, we have completed our tour of how to 
use the stationary-state
approach to find both the interdot channel conductance
$g$ and the fractional peak splitting $f$.
In order to progress further,
we must adopt a model for the barrier that gives the
energy dependence of the elements of $\delta \hat{n}$
(recall Eqs.~\ref{eq:nparts2} and \ref{eq:ncoeff2}).

\section{Peak Splitting and Conductance for a Parabolic
Barrier}

We assume that the barrier in the interdot tunneling
channel can be reasonably modeled by a parabolic one.
For an energy barrier with peak height 
$V_{0}$, such a model is plausible when
$V_{0} \simeq E_{\text{F}} \gg U$, which is the
regime of experimental interest.~\cite{Matveev5}  
The formula for a
parabolic potential $V(x)$ centered at the origin 
with half width $\xi$ is the following:
\begin{equation}
V(x)  =  \left\{ \begin{array}{ll}
      V_{0} \left( 1  - \frac{x^2}{2 \xi^2} \right) 
              & \mbox{if $|x| < \sqrt{2}\xi$} \\
      0 & \mbox{otherwise.}
                 \end{array}  \right.
\label{eq:barr1}
\end{equation}
A crucial energy scale for this barrier is the 
harmonic oscillator frequency $\omega$ of the inverted 
parabolic well.  This frequency is given by the formula
\begin{equation}
\hbar \omega = \left(\frac{1}{\pi \sqrt{2}} \right)
   \frac{(2 \pi \hbar)^2 }{2 m \lambda_{V} \xi} \, ,
\label{eq:omega}
\end{equation}
where $2 \pi/\lambda_{V}  = \sqrt{2 m V_{0}/\hbar^2}$
and $m$ is the effective mass of the electron.

The problem of transmission through and reflection from a
parabolic barrier is well known and exactly 
solvable.~\cite{ParBar,Connor}
The solutions are parabolic cylinder functions,~\cite{Abramowitz}
and the dimensionless channel conductance
is given by~\cite{Connor,Kemble}
\begin{equation}
g = \frac{1}{1 + e^{- 2 \pi y(E_{\text{F}})}} \, ,
\label{eq:gparab}
\end{equation}
where $E_{\text{F}}$ is the Fermi energy and
\begin{equation}
y(E) = \frac{E - V_{0}}{\hbar \omega} \, .
\label{eq:yofE}
\end{equation}
From these equations, it follows that
\begin{equation}
\frac{(V_{0} - E_{\text{F}})}{\hbar \omega}  =  \frac{1}{2 \pi} 
                    \ln \left( \frac{1-g}{g} \right) \, ,
\label{eq:energy&g}
\end{equation}
and, for $g \ll 1$, 
\begin{equation}
\frac{(V_{0} - E_{\text{F}})}{V_{0}} \simeq \frac{1}{2 \pi^2 \sqrt{2}} 
                   \left( \frac{\lambda_{\text{V}}}{\xi} \right)
                   |\ln g| \, .
\label{eq:EFandV}
\end{equation}
Eq.~\ref{eq:EFandV} tells us that,
even for experimental systems~\cite{Waugh,Crouch}
in which $\xi$ is quite small  
($\xi \simeq \lambda_{\text{F}}$), $E_{\text{F}}$ is close to
$V_{0}$ for $|\ln g| \ll 2\pi^2 \sqrt{2}$.
Thus, the assumption of a parabolic barrier appears 
reasonable for any measurable value of the interdot conductance.

We now consider the sizes of the various energies that appear
in our peak splitting calculations.
Equation \ref{eq:gparab} indicates that the energy scale $W$
for the variation of transmission probabilities 
is $\hbar \omega/2 \pi$.  Recalling our discussion in
Sec.~I, we have
\begin{equation}
\frac{U}{W} = \frac{2 \pi U}{\hbar \omega} \, .
\label{eq:UoverW}
\end{equation}
As observed in previous work,~\cite{Crouch,Golden1}
for symmetric dots, $U$ equals $e^{2}/(C_{\Sigma} + 2 C_{\text{int}})$,
where $C_{\Sigma}$ is the total capacitance of one of the
two dots and $C_{\text{int}}$ is the interdot capacitance.
The energy scale $\hbar \omega$ is, by comparison, only roughly
known.  From the fact that the barrier height $V_{0}$ is approximately
equal to $E_{\text{F}}$, we know that
$\lambda_{V} \simeq \lambda_{\text{F}}$.
For $\xi$, we can use the ``device resolution'' $d$, which is the
distance between the surface metallic gates and the two-dimensional
electron gas (2DEG) and is typically on the order of
$100~\text{nm}$.  The fact that the approximation $\xi \simeq d$ 
should be accurate safely within a factor of $2$
can be surmised from
calculations such as that of Davies and Nixon~\cite{Davbar} in
which they show that the potential profile induced in a 2DEG
by a narrow line gate has a half width at half maximum that 
is approximately equal to $d$.~\cite{Commbar1,Commbar2}  In the 
AlGaAs/GaAs heterostructures 
of Waugh {\it et al.}, Crouch {\it et al.}, and Livermore 
{\it et al.},~\cite{Waugh,Crouch,Livermore} where
$d$ is fairly small, about $50$ nm (approximately one Fermi 
wavelength),
further circumstantial evidence for $\xi \simeq d$ comes
from the fact that the space between the gates that form the 
interdot barrier is about $100$ nm (see Ref.~5).
It follows that, for these experimental systems,
$\hbar \omega$ is approximately $0.2 E_{\text{F}}$.  On the other hand,
$U$ is about $0.03 E_{\text{F}}$, and, therefore, to within a factor
of $2$,
$2 \pi U/\hbar \omega \simeq 1$.  For different systems in which
the Fermi wavelength is still about $50$ nm but
the gates are further from the 2DEG,~\cite{Vaart} the ratio 
$2\pi U/\hbar \omega$ is presumably even larger.  Consequently,
we expect it to be quite generally true that the
ratio $U/W = 2\pi U/\hbar \omega$ is greater than or
approximately equal to $1$.

On the other hand, since, in the sorts of experimental
situations with which we are primarily 
concerned,~\cite{Waugh,Crouch,Livermore} both $W/E_{\text{F}}$ and 
$U/E_{\text{F}}$ are much less than $1$, we 
are justified in linearizing the
single-particle energy spectrum about the Fermi surface, taking
$E(j) = E_{\text{F}} + \hbar v_{\text{F}} [k(j) - k_{\text{F}}]$,
where $k(j) = \sqrt{2 m E(j)/\hbar^2}$.

We must now calculate
$\left| B(j_{2},\alpha_{2};j_{1},\alpha_{1}) \right|$ when
$j_{1} \neq j_{2}$.  We avail ourselves 
of the exact, real solutions for the wavefunctions 
$\phi_{P j}(x)$ in the presence of a parabolic
potential~\cite{Connor,Abramowitz} ($P$ is the parity index,
which we set equal to $0$ for symmetric wavefunctions and 
$1$ for antisymmetric wavefunctions).
Connecting these to the corresponding sinusoids, we find 
that, for $x > \sqrt{2}\xi$ and $L_{\text{dot}} \gg \xi$, 
the eigenfunctions are approximately given by
\begin{equation}
\phi_{P j}(x)  =  \frac{(-1)^{P}}{\sqrt{L_{\text{dot}}}}
         \cos[k(y_{P j}) (x - \sqrt{2} \xi) + \gamma_{P}(y_{P j}) ],
\label{eq:asymp}
\end{equation}
where $y_{P j} = y(E_{P j})$ and
$k(y_{P j}) = \sqrt{2 m E_{P}(j)/\hbar^2}$.
The hard-wall boundary condition then demands that there be
an integer $n$ such that the quantity in brackets equals
$(2 n + 1)\pi/2$ when $x = L_{\text{dot}}$.

As for the phase $\gamma_{P}(y)$ itself, it can be written 
in the following general form:
\begin{equation}
\gamma_{P}(y) = (-1)^{P} R(y) + D(y) \, ,
\label{eq:gam1}
\end{equation}
If the connection to the sinusoids is made using the leading
large-$x$ forms for the parabolic cylinder 
functions,~\cite{Connor,Abramowitz} $R(y)$ and $D(y)$
are given by
\begin{eqnarray}
R(y) & = & \frac{1}{2} \arctan (e^{\pi y}) \, , \nonumber \\
D(y) & = & \frac{1}{2} [ \arg \Gamma (1/2 - i y) 
                + y \ln (4\pi\sqrt{2} \xi/\lambda_{V}) ] + D_{0} \, ,
\label{eq:gam2}
\end{eqnarray}
where $D_{0}$ is independent of $y$.

Returning to Eq.~\ref{eq:ncoeff2}, we find that, if
we restrict the integral to $x > \sqrt{2} \xi$,
we have 
\begin{eqnarray}
B' & \simeq & (-1)^{\alpha + 1} \, 
   \frac{ \sin [ D(y_2) - D(y_1) ] \,
                 \cos [ R(y_2)  + R(y_1)]} 
     {2 (k_2 - k_1) L_{\text{dot}}} \, ,
   \nonumber \\
\bar{B}' & \simeq & (-1)^{\alpha + 1} \,
     \frac{ \cos [ D(y_2) - D(y_1) ] \,
              \sin [ R(y_2)  + R(y_1)]}
     {2 (k_2 - k_1) L_{\text{dot}}} \, ,
\label{eq:Bprime}
\end{eqnarray}
where 
the bar of $\bar{B}'$ indicates that this is the term
that moves electrons from dot $\alpha$ to dot 
$\bar{\alpha}$ and we have anticipated a continuum
limit in replacing $y_{P j}$ and $k(y_{P j})$ by 
$y_{j} = y[E(j)]$ and $k_{i} = k(y_{j})$.     

In the calculation of $\bar{B}'$ and $B'$, we have 
neglected the integration
over the region $|x| < \sqrt{2}\xi$.  The results
can therefore be expected to involve errors of order
$\xi/L_{\text{dot}}$ when compared to the actual values
of $B$ and $\bar{B}$.  For $\bar{B}'$, this
is not too much of a concern since, 
when both $k_{2}$ and $k_{1}$ approach the Fermi energy,
the numerator of $\bar{B}'$ goes to $\sqrt{g}$ and
the denominator goes to zero.  Thus, for non-infinitesimal
$g$, if we restrict our wave-vectors to 
a range about the Fermi surface such that
$|k_{i} - k_{F}| \ll 1/\xi$ (in which case 
$\cos [ D(y_2) - D(y_1) ]$ can be approximated by $1$),
corrections to $\bar{B}'$ should be relatively small.  

In contrast, the term $B'$ is a bit more
problematic, for its numerator goes to zero as 
$(k_{2} - k_{1})\xi$.  Consequently, near the Fermi surface
this term is of the order of the error, and to
obtain a reliable result
we must complete the integral numerically, using the
parabolic cylinder functions in place of our sinusoids
when $|x| < \sqrt{2}\xi$.  We then find that the
form for $B'$ approximates the magnitude of 
the actual value of $B$ if, after approximating
$\cos [ R(y_2)  + R(y_1)]$ by $1$, we replace 
$[D(y_2) - D(y_1)]$ with $\kappa (y_2 - y_1)$, where 
$\kappa \simeq 0.1$ for $g \sim 0.1$ and 
$\kappa \rightarrow 0$ as $g \rightarrow 0$.
Our conclusion is that
\begin{eqnarray}
B & \simeq & (-1)^{\alpha + 1} \,
   \frac{\sin [\kappa (y_2 - y_1)]}
     {2 (k_2 - k_1) L_{\text{dot}}} \, ,  \nonumber \\
\bar{B} & \simeq & (-1)^{\alpha + 1} \, 
   \frac{\sin [ R(y_2)  + R(y_1)]}
         {2 (k_2 - k_1) L_{\text{dot}}} \, .
\label{eq:ncoeff3}
\end{eqnarray}

We can now calculate the leading parts of the energy shift 
$\Delta (\rho)$.  The contribution from hopping between
states on the same dot is given approximately by
\begin{eqnarray}
\Delta_{C}^{(2)}(\rho)  & \simeq  & -\left( \frac{U \rho^2}{4} \right)
	\frac{N_{\text{ch}}}{\pi^3} 	
	\left( \frac{2\pi U}{\hbar \omega} \right)
  \nonumber \\
 & & \hspace{0.3in} \times 
       \int_{-Y_1}^{0} \! 
                dy_{1} \! 
       \int_{0}^{Y_2} \! 
		dy_{2} \,
     \frac{\sin^2 [ \kappa (y_{2} - y_{1}) ] }
     {(y_{2} - y_{1})^{3}}  \,  ,
\label{eq:Del2C}
\end{eqnarray}
where the $y_i$'s are now measured relative to $y_{\text{F}}$
(i.e., $y_{i} \rightarrow y_{i} + y_{\text{F}}$).
The contribution from hopping between the dots obeys
\begin{eqnarray}
\Delta_{T}^{(2)}(\rho) & = & -\frac{N_{\text{ch}} U^{2}}{4}
       \int_{-\Lambda_{3}}^{0} \! \frac{dk_{1}}{\pi} \! 
       \int_{0}^{\Lambda_{4}} \! \frac{dk_{2}}{\pi} \,
   \nonumber \\
 &  & \hspace{0.2in} \times
     \frac{\sin^2 [ \tilde{R}(\hbar v_{\text{F}} k_{2}/\hbar\omega) 
               + \tilde{R}(\hbar v_{\text{F}} k_{1}/\hbar\omega)]}
     {(k_{2} - k_{1})^{2}}    \nonumber \\
 &  & \hspace{0.2in} \times \left\{
   \frac{(1 - \rho)^{2}}{\hbar v_{F}(k_{2} - k_{1}) + U(1-\rho)}
        \,       \right.             \nonumber \\
 &  & \hspace{0.6in} \left. +  \,  [\rho \rightarrow -\rho] \right\} \, ,
\label{eq:Del2T}
\end{eqnarray}
where $\tilde{R}(y) = R[y(E_{\text{F}}) + y]$
and the bracketed expression $\rho \rightarrow -\rho$ stands
for the quantity obtained by replacing $\rho$ by $-\rho$ in
the previous term.  In Eqs.~\ref{eq:Del2C} and \ref{eq:Del2T},
ultraviolet cutoffs $Y_{r}$ and $\Lambda_{r}$ 
have been inserted in recognition of the fact that
our formulas for the integrands break down at some 
distance $Y_{r}$ or $\Lambda_{r}$ from the Fermi surface.  

For the same-dot-hopping shift $\Delta_{C}^{(2)}$,
the presence of such cutoffs is essentially irrelevant
since we find this term to be effectively negligible 
no matter what the choice of $Y_{r}$.  In particular, even when 
the cutoffs are taken to infinity, this segment of the energy
shift produces a contribution to $\tilde{f}(\rho)$
(recall Eq.~\ref{eq:frho}) that is bounded by the following 
formula:
\begin{equation}
\tilde{f}_{C}(\rho) \lesssim \frac{N_{\text{ch}} \rho^{2}}{200} 
	\left(\frac{2\pi U}{\hbar \omega} \right) \, .
\label{eq:fc}
\end{equation}
The real contribution is perhaps substantially smaller
than the bound because
the integrand of Eq.~\ref{eq:Del2C} is systematically 
too large for the infinitesimal-transmission states that
correspond to $|y_1| \gtrsim 1$.

In any case, it is clear that the contribution
to $\tilde{f}(\rho)$ from same-dot hopping is
essentially negligible.
For $N_{\text{ch}} \lesssim 2$ and 
$(2\pi U/\hbar\omega) \ll 10$, 
$\tilde{f}_{C}(\rho)$ is extremely small and essentially constant 
in $g$.  Under such circumstances, it does not significantly
affect even the quantitative results.
When $(2\pi U/\hbar \omega) \gtrsim 10$, on the other hand,
it can be relatively large.  Nevertheless, it remains
unimportant, for in this regime we can 
only obtain qualitatively good results for the value of 
$\Delta_{T}^{(2)}(\rho)$, and, qualitatively, 
the upward shift of $\tilde{f}(\rho)$ induced by same-dot
hopping merely reinforces the effect from
hopping between the dots.

We now
consider the cutoffs $\Lambda_{r}$ and their impact upon
our understanding of the interdot-hopping result.
Since the integrand in Eq.~\ref{eq:Del2T} is 
reliably precise only when $k_1$ and $k_2$ are within 
$1/\xi$ of $k_{\text{F}}$, the ultraviolet cutoffs 
should be chosen such that $\Lambda_{r} \sim 1/\xi$.
It follows that, to capture with quantitative precision
the leading behavior of
the peak splitting for $g \ll 1$, the set of wave-vectors
within $1/\xi$ of the Fermi surface
should encompass the range of energies in which the 
quantity $\tilde{R}(E)$ is rapidly growing.  Consequently,
the set of wave-vectors must extend at least to
$k_{\text{F}} + k_0$, where 
$E(k_{\text{F}} + k_0) = V_{0}$.  From Eq.~\ref{eq:energy&g},
we see that the identity 
$k_0 = (V_{0} - E_{\text{F}})/\hbar v_{\text{F}}$ yields
\begin{equation}
k_0 \xi \simeq \frac{1}{2\pi \sqrt{2}} 
        \ln \! \left( \frac{1-g}{g} \right) \, .
\label{eq:kzero2}
\end{equation}
If we require that
$k_0 \xi \lesssim 1$, we see that, for $g \ll 1$,
we must have $|\ln g| \lesssim 2\pi \sqrt{2}$.
Thus, we have a lower bound on the values of $g$
for which our approximations are reliable.
Fortunately, the lower bound is very small, and
the requirement is only that $g \gtrsim 10^{-4}$.

We are now prepared to calculate $\Delta_{T}^{(2)}(\rho)$.  
After a switch to the dimensionless variables
$x_{r} = (-1)^{r} \hbar v_{\text{F}} k_{r}/U$, Eq.~\ref{eq:Del2T}
reduces to
\begin{eqnarray}
\Delta_{T}^{(2)}(\rho) & \cong & -\frac{N_{\text{ch}} U}{4 \pi^2}
       \int_{0}^{\bar{\chi}_{1}} \! dx_{1} \! 
       \int_{0}^{\bar{\chi}_{2}} \! dx_{2} \,  
    \frac{ \tilde{T}(x_1,x_2) }
         {x_{2} + x_{1} + 1 - \rho}
     \mbox{} \, 
  \nonumber \\
  & & \hspace{0.2in} +  \,  [\rho \rightarrow -\rho] \, ,
\label{eq:Del3}
\end{eqnarray} 
where $\bar{\chi}_{r} = \hbar v_{\text{F}} \Lambda_{r}/U$,
the symbol $\, \cong \,$ signifies equality modulo
terms that are independent of $\rho$, and the quantity
$\tilde{T}(x_1,x_2)$ is given by
\begin{equation}
\tilde{T}(x_{1}, x_{2}) = 
       \sin^{2} [ \tilde{R}(U x_{2}/\hbar \omega)
                 + \tilde{R}(- U x_{1}/\hbar \omega) ] \, .
\label{eq:tilT}
\end{equation}

To obtain a result with negligible
dependence on the cutoffs $\bar{\chi}_{r}$, we must have
$\bar{\chi}_{r} \gg 1$.  On the other hand, to ensure that
the answer is quantitatively reliable, we need
$\Lambda_{r} \lesssim 1/\xi$ or, equivalently, 
$\bar{\chi}_{r} \lesssim \hbar v_{\text{F}}/U \xi$.
Thus, as promised, we can only expect 
Eq.~\ref{eq:Del3} to give quantitatively reliable
results for $U \ll \hbar v_{\text{F}}/\xi$; i.e., for
$2\pi U/\hbar \omega \ll 2\pi \sqrt{2}$, where
$2\pi \sqrt{2} \simeq 9$.

Having dealt with the issue of the ultraviolet cutoffs,
we can now go about the business of evaluating the
right side of Eq.~\ref{eq:Del3}.
From the identity $\tilde{T}(0,0) = g$, it follows that
the limit $2 \pi U/\hbar \omega \rightarrow 0$ yields
the zero-width ($\xi = 0$) linear-in-$g$ equation for 
$\Delta^{(2)}(\rho)$ that was previously
derived via a transfer-Hamiltonian
approach.~\cite{Golden1,Golden2,Matveev3,Matveev4}
In contrast, in the limit 
$2\pi U/\hbar \omega \rightarrow \infty$, the energy 
shift given by Eq.~\ref{eq:Del3} is independent of the
interdot conductance for $g$ a finite distance from
both $0$ and $1$.  
The constancy of the shift follows from the fact that,
except when $g$ equals $0$ or $1$,
$\tilde{T}(x_1,x_2)$ is always $0.5$
within the bounds of integration, and the relevant
parts of the energy shift are therefore the same as for
$(2\pi U/\hbar \omega) = 0$ and
$g = 0.5$.  It should be re-emphasized, however, that
such a result for the limit 
$2\pi U/\hbar \omega \rightarrow \infty$
can only be expected to be qualitatively correct.  

What happens when the barrier width 
$\xi$ is between $0$ and $\infty$?
By performing two partial integrations of the righthand 
side of Eq.~\ref{eq:Del3} and dropping terms that
go to zero as the cutoffs $\Lambda_{r}$ become infinite, 
we find that
\begin{eqnarray}
\tilde{f}^{(1)}(\rho) & = & 
   \frac{N_{\text{ch}} g }{\pi^2} \, (1-\rho)\ln (1-\rho)
      \nonumber \\
&  & \hspace{0.1in} \mbox{} + \frac{N_{\text{ch}}}{\pi^2} 
     \left\{    \int_{0}^{\bar{\chi}_{1}} \! dx_{1} \! \left[
    \frac{\partial \tilde{T}(x_{1},0)}{\partial x_{1}}
       \,  h(\rho, x_1, 0)
             \right]   \right. 
      \nonumber \\       
&  & \hspace{0.1in} 
     \mbox{} + 
     \int_{0}^{\bar{\chi}_{2}} \! dx_{2} \! \left[
    \frac{\partial \tilde{T}(0,x_{2})}{\partial x_{2}}
       \,  h(\rho, 0, x_2)
             \right] 
      \nonumber \\
&  & \hspace{0.1in} \mbox{} \left. + 
       \int_{0}^{\bar{\chi}_{1}} \! dx_{1} \! 
       \int_{0}^{\bar{\chi}_{2}} \! dx_{2} \,  
    \frac{\partial^{2} \tilde{T}(x_{1},x_{2})}
            {\partial x_{1} \partial x_{2}}
      \,  h(\rho, x_1, x_2)   \right\}
              \nonumber \\
  & &  \mbox{} \, + \, [\rho \rightarrow -\rho] \, ,
\label{eq:frho2A}
\end{eqnarray} 
where $h(\rho,x_1,x_2) =  (x_{2} + x_{1} + 1 - \rho) 
                       \ln (x_{2} + x_{1} + 1 - \rho)$.
The first term on the righthand side of Eq.~\ref{eq:frho2A}
is the zero-width result.  The other terms, which 
go to zero in the limit $\xi \rightarrow 0$,
are the corrections due to a nonzero width.  

Numerical evaluations of Eq.~\ref{eq:frho2A} are plotted
in Fig.~2(a) for several values of the parameter 
$2\pi U/\hbar \omega$ in addition to the analytically
derived results for the limits of zero-width and 
infinite-width barriers.  A curious feature of these curves
is that the corrections to the zero-width behavior are
antisymmetric about $g = 0.5$, a property that can be
demonstrated by considering what happens to the integrands under
the transformations
$g \! \leftrightarrow \! (1-g)$ and $x_1 \! \leftrightarrow \! x_2$.
Though the antisymmetry is 
suggestive, it must be remembered that $\tilde{f}^{(1)}(\rho)$ is 
only the leading term in a perturbative expansion about
$g = 0$.  The small positive contribution
to $\tilde{f}(\rho)$ that comes from the formula for the same-dot-hopping
shift $\Delta_{C}^{(2)}$ breaks this antisymmetry,
and other higher-order corrections 
are likely to do the same.  Nevertheless, some sort of rough
antisymmetry about $g = 0.5$ is probably preserved, for, just
as we find that, at small $g$, $\tilde{f}(\rho)$ is enhanced by hopping 
connections to states with large transmission amplitudes, 
so we can expect that, at large $g$, $\tilde{f}(\rho)$ is diminished
by the fact that many of the 
occupied states from which one hops have transmission 
probabilities that are less than $g$.

Such musings aside,
we can gain further insight into the nature of our result for 
$\tilde{f}^{(1)}(\rho)$ by making a rough analytic approximation to
the righthand side of Eq.~\ref{eq:frho2A}.  To do this, it
is best to return to Eqs.~\ref{eq:frho} and \ref{eq:Del3}
and to derive the equivalent expression
\begin{eqnarray}
\tilde{f}^{(1)}(\rho) & = & \frac{ 2 N_{\text{ch}} \rho^2}{\pi^2} 
       \int_{0}^{\bar{\chi}_{1}} \! dx_{1} \! 
       \int_{0}^{\bar{\chi}_{2}} \! dx_{2}   
	 \frac{ \tilde{T}(x_1,x_2)}{(x_{2} + x_{1} + 1)}
 \nonumber \\
  & & \hspace{0.2in} \times 
	\frac{1}{(x_{2} + x_{1} + 1 - \rho)(x_{2} + x_{1} + 1+\rho)}.
\label{eq:frho2}
\end{eqnarray}

\begin{minipage}{3.27truein}
  \begin{figure}[H]
    \begin{center}
      \leavevmode
      \epsfxsize=3.27truein
      \epsfbox{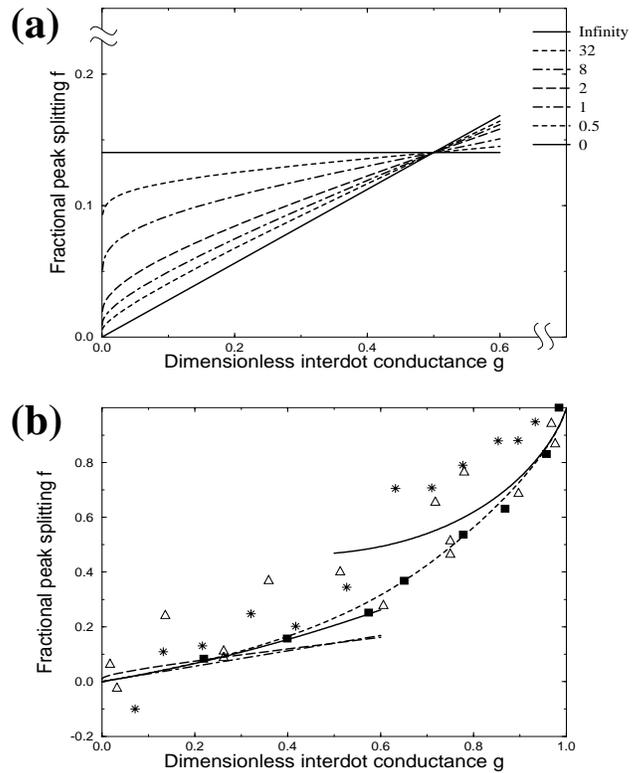}
    \end{center}
    \caption{(a) Plots of the leading $g \rightarrow 0$ term of
$f$, the fractional peak splitting, as a function of $g$,
the dimensionless interdot channel conductance.
Each curve corresponds to a different value of the
quantity $2\pi U/\hbar \omega$ (see legend on right).
All curves are for the case of two interdot tunneling
channels, $N_{\text{ch}} = 2$.  
The upward sloping solid line is the linear-in-$g$ result 
that comes from considering an interdot barrier of 
effectively zero width ($2\pi U/\hbar \omega = 0$).  
The dashed and dot-dashed curves show the
$f$-versus-$g$ dependence for finite-width 
barriers with $2\pi U/\hbar \omega$ taking on values from
$0.5$ to $32$.  The horizontal solid line gives the
leading term in the fractional peak splitting for an
infinite-width barrier ($2\pi U/\hbar \omega \rightarrow \infty$).  
The curves can only be expected to be 
quantitatively accurate when $2\pi U/\hbar \omega \ll 10$.
(b) $f$-versus-$g$ results for the full domain of $g$
when $N_{\text{ch}} = 2$.
The solid lines are the complete zero-width results in the
weak and strong-coupling limits.  These results
contain both leading and subleading terms.$^6$
The plot for the leading zero-width term in the 
weak-coupling limit ($g \rightarrow 0$) is included
as a dot-dashed curve.  The small-dashed curve that extends
from $(g,f) = (0,0)$ to $(g,f) = (1,1)$ 
is an interpolating curve that is derived from the
zero-width results.  The long-dashed line is the 
$2\pi U/\hbar \omega = 1$ curve from Fig.~2(a).  The stars,
triangles, and squares symbolize different sets of 
experimental data,$^{2,4}$ the squares being the 
most recent.$^4$}
    \label{fig:plots}
  \end{figure}
  \smallskip
\end{minipage}

\noindent We then postulate that, for small $g$, the magnitude of
the $\tilde{f}^{(1)}(\rho)$ is largely determined by
the portion of the integral that corresponds 
to $x_{2} \geq x_{0}$, where $x_{0} = \hbar v_{\text{F}} k_{0}/U$  
(recall $k_{0}$ from Eq.~\ref{eq:kzero2}).  For $x_{2}$ in this
range, $\tilde{T}(x_1,x_2)$ is on the order of $1$ and
therefore much larger than $\tilde{T}(0,0)$ when $g \ll 1$.
We label this {\it high-energy portion}
of the double integral as $\tilde{f}^{(1)}_{\text{hep}}(\rho)$.
Since, in this part of the integral, $\tilde{T}(x_1,x_2)$
varies relatively slowly between $0.15$ and $0.5$, 
we approximate it by a constant
$C_{\tilde{T}}$, where we take $C_{\tilde{T}} = 0.25$.
For $x_{0} \gtrsim 1$, we can 
drop the $\rho$'s that appear in the 
integrand of Eq.~\ref{eq:frho2}.
We then obtain
\begin{equation}
\tilde{f}^{(1)}_{\text{hep}}(\rho) \sim 
    \frac{ N_{\text{ch}} \rho^2}{4 \pi^2 (x_{0} + 1)} \, . 
\label{eq:fheprho}
\end{equation}
From the identities $f = \tilde{f}(1)$ and
$x_{0} = (\hbar\omega/2\pi U) \ln [(1-g)/g]$, we conclude
that
\begin{equation}
f^{(1)}_{\text{hep}} \sim 
    \left( \frac{ N_{\text{ch}}}{4 \pi^2} \right) 
         \frac{2\pi U/\hbar \omega}{|\ln g| + 2\pi U/\hbar \omega} \, .
\label{eq:fhep}
\end{equation}

This rough approximation to the
leading behavior of the fractional peak splitting
is only valid when $x_{0} \gtrsim 1$ and 
$k_{0} \lesssim 1/\xi$.  The condition on $x_{0}$ is
necessary to justify dropping the $\rho$'s from
the integrand in 
the high-energy portion of the double integral.
The condition on $k_{0}$ validates the approximate 
values for the magnitudes of $|A(j_2,\alpha_2;j_1,\alpha_1)|$ 
that we use throughout our calculation.
The two
restrictions together mean that the range of reliability
for our approximation to
$\tilde{f}^{(1)}_{\text{hep}}$ is given by
\begin{equation}
\frac{2 \pi U}{\hbar \omega} \lesssim |\ln g| \lesssim 2 \pi \sqrt{2}.
\label{eq:bounds}
\end{equation}
For $2 \pi U/\hbar \omega \simeq 1$, our approximations are good 
for $g$ between a couple tenths and a few ten-thousandths.

It is instructive to compare our result for 
$f^{(1)}_{\text{hep}}$ with the zero-width fractional peak 
splitting, $f^{(1)}_{\xi = 0}$.  From Eqs.~\ref{eq:fzero}
and \ref{eq:fhep}, we see that,
for $2 \pi U/\hbar \omega = 1$, the ratio 
$f^{(1)}_{\text{hep}}/f^{(1)}_{\xi = 0}$ is about
$0.6$ when $g = 0.1$ and about $25$ when $g = 0.001$.
For very weak coupling ($g \ll 0.1$), the correction
to the $\xi = 0$ result is proportionately very large, and,
as $g \rightarrow 0$, it dominates the behavior of the 
peak splitting.  On the other hand, as $g$
assumes more intermediate values ($g \sim 0.1$),
the results for $\xi = 0$ and $\xi \neq 0$ converge.

A direct comparison of our results for $f^{(1)}_{\text{hep}}$
with the full numerical results for $f^{(1)}$, which are
plotted in Fig.~2(a), 
confirms that $f^{(1)}_{\text{hep}}$ does indeed capture the 
essential $f$-versus-$g$ behavior, 
particularly as $2\pi U/\hbar \omega$
becomes larger and the exponential enhancement of the
tunneling amplitudes becomes more important.
The sharp increase in slope as $g \rightarrow 0$ can
now be understood as resulting from the fact that, in 
this limit, the high-energy portion of the peak splitting
is proportional to $(2\pi U/\hbar \omega)/|\ln g|$.
This proportionality also explains why the increase in slope
as $g \rightarrow 0$ becomes less dramatic as
$2\pi U/\hbar \omega$ decreases.
The success of our rough analytic approximation supports
the supposition that, for small $g$, a substantial
portion of the peak splitting comes from tunneling into
virtual states lying near or above the top of the barrier.

Turning to Fig.~2(b), we now examine the significance of 
the calculated finite-width corrections in the context of 
what we know about the $f$-versus-$g$ curve in the entire 
range from $g = 0$ to
$g = 1$, and we consider the implications of these
corrections for the relevant 
experiments.~\cite{Waugh,Crouch,Livermore}
The long-dashed curve in Fig.~2(b) is
the curve from Fig.~2(a) for the value 
$2\pi U/\hbar \omega = 1$, which we believe to be
appropriate for the cited experiments.  The dot-dashed line
is the leading-order-in-$g$, zero-width curve, which also
appears in Fig.~2(a).  The small-dashed curve in Fig.~2(b) is
an interpolation for the entire zero-width $f$-versus-$g$
curve.  This interpolation has been designed to match both the
second-order-in-$g$ calculation of the fractional peak
splitting for weak coupling ($g \simeq 0$) and also the 
two-term calculation for strong coupling ($g \simeq 1$), 
which were obtained in Ref.~6 and are shown as
solid curves in Fig.~2(b).  The stars, triangles, and 
squares represent different sets of experimental data. 

For the particular value of $2\pi U/\hbar \omega$ that is 
illustrated in Fig.~2(b), we see that, although the
finite-width correction to $f$
changes the answer by a large factor in the region of small $g$,
the correction is small on an absolute scale.  The difference
between the dashed curve and the dot-dashed curve never
exceeds 0.02 and therefore causes only a small correction to
the overall shape of the $f$-versus-$g$ curve.  Qualitatively,
the correction due to the finite thickness of the barrier
is quite similar to adding a small constant to $f$
near $g = 0$ and then decreasing the slope of the $f$-versus-$g$
curve at small $g$.  This qualitative similarity follows from
the fact that the region where $f$ drops rapidly to zero, at
very small $g$, is almost invisible in the plot.  Consequently,
the correction to the zero-width curve might be hard to 
distinguish from the effects of a small interdot capacitance,
which have already been included in analyzing the data.
We therefore conclude that introduction of
the finite thickness correction has little effect on the
agreement between theory and the existing experimental data,
for which $2\pi U/\hbar \omega \simeq 1$.  Nevertheless,
such corrections may be important in future experiments.

\section{Conclusion}

By developing a new approach to the coupled-dot problem that
relies upon the non-interacting, single-particle eigenstates
of the full coupled-dot system, 
we solve for the leading correction to zero-width,
weak-coupling results that were derived in previous 
work.~\cite{Golden1,Golden2,Matveev3,Matveev4}  The 
nonzero barrier width $\xi$ and finite barrier height
$V_{0}$ mean that the off-diagonal ``hopping terms'' vary 
exponentially with the energies of the states they connect.  
For a small interdot channel conductance ($g \ll 1$), the 
resulting enhancement of tunneling to ``high-energy'' 
states above the barrier leads to an 
increase in the magnitude of the fractional peak splitting 
$f$ observed at a given value of $g$.  For a parabolic barrier, the
magnitude of this increase grows with the value of the
ratio $2 \pi U/\hbar \omega$, where $U$ is the interdot
charging energy and $\omega$ is the frequency of the 
inverted parabolic well.  Except in a very small region near 
$g = 0$ where $f$ behaves like $(2\pi U/\hbar \omega)/|\ln g|$,
the increase in $f$ is accompanied by a decrease in the
slope of the $f$-versus-$g$ curve.  The effect upon the
overall shape of the $f$-versus-$g$ curve is not very
substantial for experiments in which 
$(2\pi U/\hbar \omega) \simeq 1$ but could be crucial
in interpreting experiments involving significantly
wider barriers.  

One might worry that the finite-width corrections
to higher-order terms in the weak-coupling expansion
could lead to a more dramatic alteration of the 
$f$-versus-$g$ curve.  However, the corrections to such 
``large-$g$'' terms should be muted by the fact that, as 
$g$ increases, there is less difference between tunneling 
amplitudes between states at the Fermi energy and tunneling 
amplitudes between a state at the Fermi energy and a state lying 
above the barrier.

A more vital source of concern might be the
treatment of the electron-electron interactions in the vicinity
of the barrier.  Clearly, the use of a sharp step function
in the equation for $\hat{n}$ (recall Eq.~\ref{eq:nstep}) is an 
artifice.  A more realistic model would account for the
fact that, though electrons in and about the interdot channel
still repel one another locally, their interactions with the
rest of the electrons in the system are screened
by the surface gates. 

Finally, one might wonder whether higher-order corrections to $f$
preserve at least a rough antisymmetry about $g \simeq 0.5$.
We have seen that the leading small-$g$ correction, when directly
extended to $g = 1$, changes sign and becomes negative for
$g > 0.5$.  Although a proper calculation of the behavior at
such large values of $g$ requires consideration of higher-order
diagrams, which we have neglected, we believe that the negativity
of the correction to $f$ at large values of $g$ is a generally
right physical feature.  When $g$ is large and the reflection
probability at the Fermi energy is therefore small, 
the energy dependence of the reflection amplitude, for
$\xi \neq 0$, should lead to a decrease in $f$ as a result of
the enhanced reflection coefficient for occupied states
lying below the barrier.    


\section*{Acknowledgments}
The authors thank C.~Livermore, C.~H.~Crouch, and
R.~M.~Westervelt 
for helpful discussions.  This work was supported by the NSF 
through the Harvard Materials Research Science and
Engineering Center, Grant No. DMR94-00396.

\end{multicols}


\begin{references}
\bibitem{CommSmear} Use of a sharp step function is
somewhat unrealistic.  However, a ``smeared-out''
step function such as $[1 + \tanh(x/\xi)]/2$ creates a 
region of characteristic width $\xi$ about $x = 0$ where
electrons are neither in dot 1 nor dot 2 from the point
of view of $\hat{n}$ (recall Eq.~\ref{eq:nstep}).
Under such circumstances, the value of $\hat{n}$ can
be changed simply by having a net movement of electrons
toward the barrier---without any tunneling between the
dots.  Clearly, such a picture misses the crucial
fact that electrons in the vicinity of the barrier do, in 
fact, repel one another.  Until a better model
of the electron repulsion in the barrier region is 
developed, it is best to stick with a sharp step and
thereby avoid inventing a region near the barrier in
which electrons do not interact.
\bibitem{Commbar1} For a zero-width gate, the half width
equals $d$.  For broader gates, the half width is larger,
but it grows relatively slowly 
with the gate width $w$, only reaching about $1.4 d$ for
$w = 2 d$.  For much larger values of $w$, the potential 
profile becomes rather flat beneath the gate,
and a parabolic model is not useful.
\bibitem{Commbar2} Of course, in considering tunneling 
between the dots, we are not dealing with a line gate but
two narrow split gates separated by a gap of length $\ell$.
However, for weakly coupled dots, the
potential profile in the gap is presumably reasonably well
approximated by that of a line gate with an effective 
width $w_{\text{eff}}$.  This 
should remain true even when electron-electron interactions 
are included because, near the barrier peak and for 
$g \simeq 0$, the corrections due
to electronic screening should be relatively small.
In any case, these corrections
tend to counteract the broadening due to
$w_{\text{eff}} \neq 0$.
\end{references}
\end{document}